\documentclass[12pt]{article}
\usepackage{amssymb}
\newcommand{\be}{\begin{eqnarray}}
\newcommand{\ee}{\end{eqnarray}}
\newcommand{\non}{\nonumber}
\newcommand{\g}{\ensuremath{\mathsf{g}}}

\begin{document}

\begin{titlepage}
\strut\hfill UMTG--229
\vspace{.5in}
\begin{center}

\LARGE The boundary $N=2$ supersymmetric sine-Gordon model\\[1.0in]
\large Rafael I. Nepomechie\\[0.8in]
\large Physics Department, P.O. Box 248046, University of Miami\\[0.2in]  
\large Coral Gables, FL 33124 USA\\

\end{center}

\vspace{.5in}

\begin{abstract}
We construct an action for the $N=2$ supersymmetric sine-Gordon model
on the half-line, which we argue is both supersymmetric and integrable.  
The boundary interaction depends on three continuous boundary parameters, 
as well as the bulk mass parameter.
\end{abstract}

\end{titlepage}

\setcounter{footnote}{0}

\section{Introduction}\label{sec:intro}

Many important results have been obtained for two-dimensional bulk
$N=2$ superconformal field theories \cite{SCFT} and their integrable
perturbations \cite{pertbs}.  Some of these results have already
played an important role in closed string theory.  (See, e.g.,
\cite{We, reviews, Po}, and references therein.)  Soon after the
pioneering work of Ghoshal and Zamolodchikov \cite{GZ} on boundary
integrable field theories, Warner \cite{Wa} initiated a program to
generalize the bulk $N=2$ results to the boundary case, which is
evidently relevant for open strings.

We consider here the $N=2$ supersymmetric sine-Gordon (SSG) model
\cite{KU} on the half-line $x \le 0$, which is the prototypical
boundary integrable model with extended supersymmetry, as is the $N=0$
(ordinary) boundary sine-Gordon model in the absence of supersymmetry. 
Specifically, we construct an action which we argue is both
supersymmetric and integrable.  This action has two further notable
features.  First, the boundary interaction has three continuous
boundary parameters -- one more than in the cases of $N=0$ \cite{GZ}
or $N=1$ \cite{IOZ, Ne}.  We expect that this additional parameter
affords the possibility of interpolating continuously between
Neveu-Schwarz and Ramond boundary conditions.  Secondly, the boundary
interaction depends also on the bulk mass parameter $g$.  Our action
is exact for the bulk massless case; but in the massive case, it is
only an approximation to first order in $g$.  While the possibility of
such bulk-boundary mixing was already anticipated in \cite{GZ}, it
does not occur for $N=0$, and first appears for $N=1$.  

The outline of this Letter is as follows.  In Section 2, we present
the action of the boundary $N=2$ SSG model in terms of several unknown
functions.  We then proceed to determine these functions from the
requirements of supersymmetry and integrability.  In Section 3 we
treat the simpler massless bulk case, and then in Section 4 we
consider the more difficult massive bulk case.  We briefly discuss our
results in Section 5.  In the Appendix we construct, using results
from \cite{KU}, the bulk conserved currents which we need for our
analysis.

\section{The action}\label{sec:SSG}

The Euclidean-space action of the boundary $N=2$ supersymmetric
sine-Gordon model is given by
\be
S &=& \int_{-\infty}^{\infty} dy \int_{-\infty}^{0} dx\ {\cal L}_{0}
+ \int_{-\infty}^{\infty} dy\  {\cal L}_{b} \,,
\ee
where the bulk Lagrangian density is given by \cite{KU}
\be
{\cal L}_{0} &=&
{1\over 2}\left(-\partial_{z}\varphi^{-} \partial_{\bar z} \varphi^{+}
-\partial_{\bar z}\varphi^{-} \partial_{z} \varphi^{+} 
+ \bar \psi^{-} \partial_{z} \bar \psi^{+} 
+ \psi^{-} \partial_{\bar z} \psi^{+} 
+ \bar \psi^{+} \partial_{z} \bar \psi^{-} 
+ \psi^{+} \partial_{\bar z} \psi^{-} \right)  \non \\
&+& g \cos \varphi^{+} \bar \psi^{-} \psi^{-} +
g \cos \varphi^{-} \bar \psi^{+} \psi^{+}
+ g^{2} \sin \varphi^{+} \sin \varphi^{-} \,, 
\label{bulkL}
\ee
where $\varphi^{\pm}$ form a complex scalar field; $\psi^{\pm}$ and
$\bar \psi^{\pm}$ are the components of a complex Dirac Fermion field;
$g$ is the bulk mass parameter; and $z={1\over 2}(y+ix)$, $\bar
z={1\over 2}(y-ix)$.
(The standard dimensionless bulk coupling constant $\beta$ is absent,
since it can be made to appear in an overall factor ${1\over
\beta^{2}}$ in front of the action by appropriate field rescalings.  We
consider the classical limit $\beta \rightarrow 0$.)  Motivated in
part by our $N=1$ result \cite{Ne}, we propose the following boundary
Lagrangian at $x=0$:
\be
\lefteqn{
{\cal L}_{b} = -{i\over 2}(\bar \psi^{+} \psi^{-}+\bar \psi^{-} \psi^{+})
-{1\over 2} a^{-} \partial_{y} a^{+} 
- {\cal B}(\varphi^{+} \,, \varphi^{-})} \label{boundL} \\ 
&&+{1\over 2} \left( f^{+}(\varphi^{+}) a^{+} 
+ g \tilde f^{+}(\varphi^{+}) a^{-}\right) 
(\psi^{-} + \bar \psi^{-})
+ {1\over 2} \left(f^{-}(\varphi^{-}) a^{-} 
+ g \tilde f^{-}(\varphi^{-}) a^{+}\right) 
(\psi^{+} + \bar \psi^{+}) \,,  \non
\ee
where $a^{\pm}$ are Fermionic boundary degrees of freedom which
anticommute with $\psi^{\pm}$ and $\bar \psi^{\pm}$.  Such boundary
degrees of freedom were introduced by Ghoshal and Zamolodchikov
\cite{GZ} to describe the Ising model in a boundary magnetic field,
and were further exploited in \cite{Wa, Ne, AKL, MN}.  Moreover,
$f^{\pm}(\varphi^{\pm})$, $\tilde f^{\pm}(\varphi^{\pm})$ and ${\cal
B}(\varphi^{+} \,, \varphi^{-})$ are potentials (functions of the
scalar fields $\varphi^{\pm}$, but not of their derivatives), which
are still to be determined. Note that, as mentioned in the 
Introduction, the boundary Lagrangian depends on the bulk mass 
parameter $g$.

Variation of the action gives the classical equations of motion. For 
the bulk, the equations are 
\be
\partial_{z}\partial_{\bar z} \varphi^{\pm} &=& -g \sin \varphi^{\mp} 
\psi^{\pm} \bar\psi^{\pm} - g^{2} \sin \varphi^{\pm} \cos \varphi^{\mp}
\,, \non \\
\partial_{z} \bar\psi^{\mp} &=& -g \cos \varphi^{\mp} \psi^{\pm} 
\,, \non \\
\partial_{\bar z} \psi^{\mp} &=& g \cos \varphi^{\mp} \bar \psi^{\pm} \,.
\label{bulkeom}
\ee
The boundary conditions at $x=0$ are
\be
\partial_{x}\varphi^{\pm} &=& - {\partial {\cal B}\over \partial 
\varphi^{\mp}} + {1\over 2} \left({\partial f^{\mp} \over \partial 
\varphi^{\mp}} a^{\mp} + g {\partial \tilde f^{\mp} \over \partial 
\varphi^{\mp}} a^{\pm} \right) (\psi^{\pm} + \bar \psi^{\pm}) 
\,, \non \\
\psi^{\pm} - \bar\psi^{\pm} &=& i f^{\pm} a^{\pm} +
i g \tilde f^{\pm} a^{\mp} \,, \non \\
\partial_{y}a^{\pm}  &=& f^{\mp} (\psi^{\pm} + \bar \psi^{\pm})
+ g \tilde f^{\pm} (\psi^{\mp} + \bar \psi^{\mp})\,.
\label{boundeom}
\ee

Following a similar strategy as in \cite{GZ,Ne}, we now proceed to
determine the functions $f^{\pm}(\varphi^{\pm})$, $\tilde
f^{\pm}(\varphi^{\pm})$ and ${\cal B}(\varphi^{+} \,, \varphi^{-})$
which appear in the boundary action by demanding both $N=2$
supersymmetry and integrability.

That the bulk $N=2$ SSG model is integrable implies that it has an
infinite number of classical integrals of motion constructed from the
densities $T_{s+1}$, $\overline{T}_{s+1}$, $\Theta_{s-1}$,
$\overline{\Theta}_{s-1}$, which obey
\be
\partial_{\bar z} T_{s+1} = \partial_{z} \Theta_{s-1} \,, 
\qquad 
\partial_{z} \overline{T}_{s+1} 
= \partial_{\bar z} \overline{\Theta}_{s-1} \,.
\label{continuity}
\ee 
The densities for $s={1\over 2}, 1, 3$ are given in the 
Appendix \ref{conserved}. 

As observed in \cite{GZ}, it follows from the continuity Eqs. 
(\ref{continuity}) that the boundary model has the integral of motion
\be
P_{s}= \int_{-\infty}^{0}dx\ (T_{s+1} + \overline{T}_{s+1} - 
\Theta_{s-1} - \overline{\Theta}_{s-1}) + i\Sigma_{s}(y) \,,
\label{iom}
\ee
provided that the following condition holds at $x=0$
\be
T_{s+1} - \overline{T}_{s+1} + \Theta_{s-1} - \overline{\Theta}_{s-1}
= \partial_{y} \Sigma_{s}(y) \,,
\label{constraint}
\ee 
where $\Sigma_{s}(y)$ is a local boundary term. Hence, our task 
reduces to investigating the constraints (\ref{constraint}) for
$s={1\over 2}$ (to have supersymmetry) and $s=3$ (to have
integrability).  Since this remains a formidable undertaking, it is
convenient to treat separately the bulk massless and bulk massive
cases.

\section{Bulk massless case}\label{sec:massless}

We begin by considering the simpler case of no bulk mass, $g=0$.
The supersymmetry constraints are
\be
T_{3\over 2}^{\pm} - {\overline{T}}_{3\over 2}^{\pm} 
= \partial_{z} \varphi^{\pm} \psi^{\pm} 
- \partial_{\bar z} \varphi^{\pm} \bar \psi^{\pm} 
= -i \partial_{x}\varphi^{\pm}(\psi^{\pm} + \bar\psi^{\pm})
+ \partial_{y}\varphi^{\pm}(\psi^{\pm} - \bar\psi^{\pm})
= \partial_{y} 
\Sigma_{1\over 2}^{\pm} \,.
\label{susyconstraint}
\ee
Let us assume that the boundary terms have the form
\be
\Sigma_{1\over 2}^{\pm} = i \g^{\pm}(\varphi^{\pm}) a^{\pm} \,.
\label{boundterm}
\ee 
We find, with the help of the boundary conditions (\ref{boundeom}),
that the constraints (\ref{susyconstraint}) are satisfied provided
that
\be
f^{\pm}(\varphi^{\pm})={\partial \g^{\pm}(\varphi^{\pm})
\over \partial \varphi^{\pm}} \,, \qquad 
\g^{\pm}(\varphi^{\pm}) f^{\mp}(\varphi^{\mp}) = 
{\partial {\cal B}\over \partial \varphi^{\mp}} \,.
\ee
These equations imply the relation
\be
f^{+}(\varphi^{+}) f^{-}(\varphi^{-}) = 
{\partial^{2} {\cal B}\over \partial \varphi^{+} 
\partial \varphi^{-}} \,.
\label{fBrelation}
\ee 

We next turn to the integrability constraint
\be
T_{4} - \overline{T}_{4} = \partial_{y} \Sigma_{3} \,.
\label{s3constraint}
\ee
Eliminating $a^{\pm}$ from the boundary conditions (\ref{boundeom}), 
we obtain
\be
\partial_{x}\varphi^{\pm} &=& - {\partial {\cal B}\over \partial 
\varphi^{\mp}} - {i\over 2} {\partial \ln f^{\mp} \over \partial 
\varphi^{\mp}} (\psi^{\mp} - \bar \psi^{\mp}) 
(\psi^{\pm} + \bar \psi^{\pm}) 
\,, \non \\
\partial_{y}\psi^{\pm} - \partial_{y}\bar\psi^{\pm} 
&=& {\partial \ln f^{\pm} \over \partial 
\varphi^{\pm}} \partial_{y} \varphi^{\pm} (\psi^{\pm} - \bar \psi^{\pm})
+i f^{+} f^{-} (\psi^{\pm} + \bar \psi^{\pm}) 
\,. 
\label{masslessboundeom}
\ee
We substitute these boundary conditions, as well as the bulk equations
of motion (\ref{bulkeom}) with $g=0$, into the LHS of the constraint
(\ref{s3constraint}).
An examination of the pure Bosonic terms reveals that this constraint
can be satisfied only if the potential ${\cal B}$ obeys
\be
{\cal B} = -4 {\partial^{2} {\cal B}\over \partial \varphi^{- 2}}
= -4 {\partial^{2} {\cal B}\over \partial \varphi^{+ 2}} \,.
\ee
We conclude that ${\cal B}$ is given by
\be
{\cal B} = \alpha \cos ({1\over 2}(\varphi^{+} - \varphi^{+}_{0}))
\cos ({1\over 2}(\varphi^{-} - \varphi^{-}_{0})) \,,
\label{Bresult}
\ee
where $\alpha$, $\varphi^{\pm}_{0}$ represent three independent real
parameters.\footnote{Since we regard the fields $\varphi^{\pm}$ as a
complex-conjugate pair, then so are $\varphi^{\pm}_{0}$.  That is,
$\varphi^{\pm}_{0}$ count as one complex, or two real, parameters.}
It follows from the relation (\ref{fBrelation}) that 
$f^{\pm}(\varphi^{\pm})$ are given by
\be
f^{\pm}(\varphi^{\pm}) = 
{\sqrt{\alpha}\over 2}\sin ({1\over 2}(\varphi^{\pm} - \varphi^{\pm}_{0}))
\,,
\label{fresult}
\ee
Remarkably, with these choices of ${\cal B}$ and $f^{\pm}$, the
remaining terms (i.e., those which are not pure Bosonic) on the LHS of
(\ref{s3constraint}) can also be expressed as a total $y$ derivative.  The
computation, which is rather arduous, follows the general lines of the
simpler $N=1$ case \cite{Ne}.  However, there is a final ``twist'',
requiring use of the amusing identity
\be
\lefteqn{
\partial_{y} \left( \psi^{+} \psi^{-} + \bar\psi^{+}\bar\psi^{-} 
- \bar\psi^{+}\psi^{-} - \psi^{+}\bar\psi^{-} \right)} \non \\
& & = (A+B) \psi^{+} \psi^{-} + (-A+B) \bar\psi^{+}\bar\psi^{-} 
- B \bar\psi^{+}\psi^{-} - B \psi^{+}\bar\psi^{-} \,,
\label{identity}
\ee
where 
\be 
A = 2 i f^{+} f^{-} \,, \qquad
B = {\partial \ln f^{+} \over \partial 
\varphi^{+}} \partial_{y} \varphi^{+} + 
{\partial \ln f^{-} \over \partial 
\varphi^{-}} \partial_{y} \varphi^{-} \,.
\ee 
 
\section{Bulk massive case}\label{sec:massive}

We now consider the general case $g \ne 0$. The supersymmetry 
constraints now read
\be
T_{3\over 2}^{\pm} - {\overline{T}}_{3\over 2}^{\pm} 
+ \Theta_{-{1\over 2}}^{\pm} - \overline{\Theta}_{-{1\over 2}}^{\pm}
&=& -i \partial_{x}\varphi^{\pm}(\psi^{\pm} + \bar\psi^{\pm})
+ \partial_{y}\varphi^{\pm}(\psi^{\pm} - \bar\psi^{\pm})
+ g \sin \varphi^{\pm} (\psi^{\mp} + \bar\psi^{\mp}) \non  \\
&=& \partial_{y} \Sigma_{1\over 2}^{\pm} \,.
\label{susyconstraintmassive}
\ee
Given the potential ${\cal B}$ (\ref{Bresult}), \footnote{As discussed
below, the potential ${\cal B}$ does not change when a bulk mass is
turned on.} the only way that we have found to satisfy this constraint
is to introduce the $g$-dependent interactions given in the boundary
conditions (\ref{boundeom}).  Indeed, this is how we first arrived at
those $g$-dependent corrections.  Making the Ansatz
\be
\Sigma_{1\over 2}^{\pm} = i \g^{\pm}(\varphi^{\pm}) a^{\pm} 
+  i g \tilde \g^{\pm}(\varphi^{\pm}) a^{\mp} \,,
\label{boundtermmassive}
\ee 
we derive the relations
\be
f^{\pm}(\varphi^{\pm}) = {\partial \g^{\pm}(\varphi^{\pm})
\over \partial \varphi^{\pm}} \,, \qquad 
\tilde f^{\pm}(\varphi^{\pm})={\partial \tilde  \g^{\pm}(\varphi^{\pm})
\over \partial \varphi^{\pm}} \,,
\label{massiverelations1}
\ee
and
\be 
\sin \varphi^{\pm} = i \left( 
\g^{\pm} \tilde f^{\pm}  + \tilde \g^{\pm} f^{\pm} \right)
\,, \qquad 
\g^{\pm} f^{\mp} + g^{2} \tilde \g^{\pm} \tilde f^{\mp} = 
{\partial {\cal B}\over \partial \varphi^{\mp}} \,.
\ee
It follows that
\be
\g^{\pm} \tilde \g^{\pm} &=& i \cos \varphi^{\pm} + C^{\pm}
\,, \label{massiverelations2}  \\
f^{+} f^{-}
+ g^{2} \tilde f^{+} \tilde f^{-}
&=& {\partial^{2} {\cal B}\over \partial \varphi^{+} 
\partial \varphi^{-}} \,,
\label{massiverelations3}
\ee 
where $C^{\pm}$ are arbitrary integration constants. 

The relations (\ref{massiverelations1}), (\ref{massiverelations2})
and (\ref{massiverelations3}) are sufficient to ensure $N=2$
supersymmetry.  Further restrictions come from implementing the
integrability constraint
\be
T_{4} - \overline{T}_{4} + \Theta_{2} - \overline{\Theta}_{2}
= \partial_{y} \Sigma_{3}(y) \,,
\label{s3massive}
\ee 
Due to the difficulty of this computation, we now restrict our analysis  
to first order in $g$. Elimination of $a^{\pm}$ from the boundary 
conditions (\ref{boundeom}) now gives (dropping terms of order
$g^{2}$ and higher)
\be
\partial_{x}\varphi^{\pm} &=&  \mbox{massless}  +
i g M^{\pm} \psi^{\pm} \bar \psi^{\pm} \,, \non \\
\partial_{y}\psi^{\pm} - \partial_{y}\bar\psi^{\pm} 
&=&  \mbox{massless}   
- g M^{\mp} \partial_{y}\varphi^{\pm} (\psi^{\mp} - \bar \psi^{\mp})
+2 i g f^{\pm} \tilde f^{\pm} (\psi^{\mp} + \bar \psi^{\mp}) \,, 
\label{massiveboundeom}
\ee
where 
\be
M^{\pm} = {\tilde f^{\mp}\over f^{\pm}}{\partial \ln f^{\mp}\over 
\partial \varphi^{\mp}} -  
{1\over f^{\pm}}{\partial \tilde f^{\mp}\over 
\partial \varphi^{\mp}} \,,
\ee 
and ``massless'' represents the corresponding $g=0$ result
(\ref{masslessboundeom}).  Examination of the pure Bosonic terms in
(\ref{s3massive}) leads to the same result (\ref{Bresult}) for the
potential ${\cal B}$.  To first order in $g$, the relation
(\ref{massiverelations3}) reduces to the corresponding massless one
(\ref{fBrelation}); hence, we are lead to the same result
(\ref{fresult}) for $f^{\pm}(\varphi^{\pm})$.  This determines 
$\g^{\pm}(\varphi^{\pm})$, up to an integration constant.  We can then
use (\ref{massiverelations2}) to determine 
$\tilde \g^{\pm}(\varphi^{\pm})$.  The choice $C^{\pm} = i \cos
\varphi^{\pm}_{0}$ gives the simple result
\be
\tilde f^{\pm}(\varphi^{\pm}) = {i\over \sqrt{\alpha}}
\sin ({1\over 2}(\varphi^{\pm} + \varphi^{\pm}_{0}))
\,.
\label{tildefresult}
\ee
These choices for $f^{\pm}$ and $\tilde f^{\pm}$ satisfy the
entire set of relations (\ref{massiverelations1}),
(\ref{massiverelations2}) and (\ref{massiverelations3}) to first order
in $g$, thereby ensuring supersymmetry to that order.  One must now
consider the contributions of order $g$ to the LHS of the constraint
(\ref{s3massive}) that are not pure Bosonic. Remarkably, after another 
long computation, we find that all such contributions can be 
expressed as a total $y$ derivative. We conclude that, to first order 
in $g$, both $P_{1\over 2}^{\pm}$ and $P_{3}$ are integrals of motion.

\section{Discussion}\label{sec:discuss}

We have demonstrated that the boundary $N=2$ SSG action (\ref{bulkL}),
(\ref{boundL}), with ${\cal B}$, $f^{\pm}$ and $\tilde f^{\pm}$ given
by Eqs.  (\ref{Bresult}), (\ref{fresult}) and (\ref{tildefresult}),
respectively, has the integrals of motion $P_{1\over 2}^{\pm}$ and
$P_{3}$ (\ref{iom}), to first order in the bulk mass parameter $g$. 
The conservation of $P_{1\over 2}^{\pm}$ means that the model has
on-shell $N=2$ supersymmetry.  The conservation of $P_{3}$ provides
strong evidence that the model is integrable.  We therefore conjecture
that the quantized model (with appropriate boundary corrections that
are higher order in $g$) has $N=2$ supersymmetry and is integrable.

Our action bears some similarity to the so-called Landau-Ginzburg
model discussed in Section 7 of \cite{Wa} with bulk superpotential $W
= \cos \varphi$.  However, there is an essential difference: the
boundary (Bosonic) potential $|W|^{2}$ proposed in \cite{Wa}, which
differs from ours (\ref{Bresult}), is not integrable.  In \cite{Wa},
the boundary potential is selected solely on the basis of
supersymmetry.  While that choice does achieve supersymmetry, it is
not the unique such choice.  Indeed, our action also maintains
supersymmetry, albeit in a more intricate manner through $g$-dependent
boundary interactions, which evidently is the price to be paid for
integrability.  As a bonus, our boundary action contains more boundary
parameters (a total of three, in addition to the bulk mass parameter)
than the boundary action considered previously \cite{Wa}.

It remains a challenge to work out corrections to the boundary action
that are higher order in $g$.  A naive guess for the case
$\varphi^{\pm}_{0}=0$ is
\be
f^{\pm}(\varphi^{\pm}) = \gamma \sin {\varphi^{\pm}\over 2} \,, \qquad 
\tilde f^{\pm}(\varphi^{\pm}) = {i\over 2\gamma} \sin {\varphi^{\pm}\over 2} 
\,, \qquad \gamma=\sqrt{{\alpha\over 8}+\sqrt{{\alpha^{2}\over 
64}+{g^{2}\over 4}}} \,.
\ee
Indeed, these expressions have the correct $g \rightarrow 0$ limit,
and satisfy the relations (\ref{massiverelations1}),
(\ref{massiverelations2}) and (\ref{massiverelations3}) which ensure
supersymmetry.  However, we have not attempted to check the
conservation of $P_{3}$ with this choice.

It would clearly be advantageous to have superfield formulations of 
the boundary $N=1, 2$ SSG models. Presumably, this would entail making 
more precise the notion of the ``boundary of a superspace''. Some 
progress in this direction has already been made in \cite{IM, Ho}.

We intend to present exact boundary scattering matrices and 
thermodynamics of the model in forthcoming publications.

\section*{Acknowledgments}

Correspondence with C. Ahn, M. Grisaru and E. Martinec is gratefully 
acknowledged.
This work was supported in part by the National Science Foundation
under Grant PHY-9870101.

\appendix

\section{Bulk conserved currents}\label{conserved}

We explicitly construct here the first few bulk classically-conserved
currents for the $N=2$ supersymmetric sine-Gordon model.  For this
task, a superfield approach is essential.  (For an exposition of $N=2$
superfield formalism in two dimensions, see Chapter 23 of \cite{We}.) 
Following Kobayashi and Uematsu \cite{KU}, we introduce the covariant
derivatives
\be
D_{\pm} = {\partial\over \partial \theta^{\pm}}+ {1\over 2} 
\theta^{\mp}\partial_{z} \,, \qquad
\overline{D}_{\pm} = {\partial\over \partial \bar\theta^{\pm}}+ {1\over 2} 
\bar\theta^{\mp}\partial_{\bar z} \,,
\ee
which obey
\be
\left\{ D_{+} \,, D_{-} \right\} &=& \partial_{z} \,, \qquad
\left\{ \overline{D}_{+} \,, \overline{D}_{-} \right\} = 
\partial_{\bar z} \,, \non \\
\left\{ D_{+} \,, \overline{D}_{\pm} \right\} &=& 0 =
\left\{ D_{-} \,, \overline{D}_{\pm} \right\} \,, \qquad
D_{\pm}^{2} = 0 = \overline{D}_{\pm}^{2} \,, \non \\
\left[ D_{\pm} \,, \partial_{z} \right] &=& 0 =
\left[ \overline{D}_{\pm} \,, \partial_{z} \right] \,,
\ee
and the chiral superfields $\phi^{\pm}$ obeying
\be
D_{\mp} \phi^{\pm} = \overline{D}_{\mp}\phi^{\pm} = 0  \,.
\ee 
The components of these superfields are given by
\be
\varphi^{\pm} &=& \phi^{\pm} \Big\vert \,, \qquad
F^{\pm} =  \overline{D}_{\pm}D_{\pm} \phi^{\pm} \Big\vert \,, \non \\
\psi^{\pm} &=& D_{\mp}\phi^{\mp} \Big\vert \,, \qquad
\bar\psi^{\pm} =  \overline{D}_{\mp} \phi^{\mp} \Big\vert \,,
\ee
where $\Big\vert$ is the standard shorthand for the projection
$\Big\vert_{\theta^{\pm}=\bar\theta^{\pm}=0}$. The bulk action is
given by \cite{KU}
\be
S = \int d^{2}z d^{4}\theta \phi^{+} \phi^{-} 
+ g \int d^{2}z d^{2}\theta^{+} \cos \phi^{+} 
+ g \int d^{2}z d^{2}\theta^{-} \cos \phi^{-} \,,
\ee
from which follow the bulk equations of motion
\be
\overline{D}_{\pm} D_{\pm} \phi^{\pm} = g \sin \phi^{\mp}  \,.
\label{superfieldeom}
\ee 
The corresponding component expressions for the action (up to total
derivatives) and equations of motion (after eliminating the auxiliary
fields $F^{\pm}$ by means of their field equations) are given by Eqs. 
(\ref{bulkL}) and (\ref{bulkeom}), respectively.

We are now in position to work out the classically-conserved currents.
With the help of the equations of motion, one can show that
the quantity $X = D_{+} \phi^{+} D_{-} \phi^{-}$ \cite{KU} obeys
\be
\overline{D}_{\pm} X = D_{\mp} A^{\pm}  \,, 
\label{superconservation}
\ee
with $A^{\pm}= \mp g \cos \phi^{\mp}$. Acting on both sides of the 
first of these equations with $D_{\mp} \overline{D}_{-}$, we obtain
\be
\partial_{\bar z}\left( D_{\mp}X \right) \Big\vert =
-\partial_{z}\left( \overline{D}_{\pm}A^{\mp} \right) \Big\vert \,.
\ee
We therefore identify the supercurrents
\be
\pm T^{\pm}_{3\over 2} = D_{\mp}X \Big\vert = 
\pm \partial_{z} \varphi^{\pm} \psi^{\pm} \,,
\qquad \mp\Theta^{\pm}_{-{1\over 2}}= \overline{D}_{\pm} A^{\mp} \Big\vert 
= \mp g \sin \varphi^{\pm} \bar \psi^{\mp} \,.
\ee
Moreover, acting on both sides of the first equation in
(\ref{superconservation}) with $D_{-} D_{+}\overline{D}_{-}$, we
obtain
\be
\partial_{\bar z}\left( D_{-}D_{+}X \right) \Big\vert =
\partial_{z}\left( \overline{D}_{-}D_{-}A^{+} \right) \Big\vert \,,
\ee
and we therefore identify the energy-momentum tensor
\be
-T_{2} &=& D_{-}D_{+}X \Big\vert = 
-\partial_{z} \varphi^{+}\partial_{z}\varphi^{-} 
+ \psi^{-} \partial_{z} \psi^{+} \,, \non \\ 
-\Theta_{0} &=&  \overline{D}_{-}D_{-}A^{+} \Big\vert =
g \cos \varphi^{-} \bar \psi^{+} \psi^{+} +
g^{2} \sin \varphi^{-} \sin \varphi^{+} \,.
\ee

Similarly, the quantity $\overline{X} 
= \overline{D}_{+} \phi^{+} \overline{D}_{-} \phi^{-}$ obeys
\be
D_{\pm} \overline{X} = \overline{D}_{\mp} \overline{A}^{\pm} \,, 
\ee
with $\overline{A}^{\pm}=-A^{\pm}$. It follows that
\be
\partial_{z}\left(\overline{D}_{\mp}\overline{X} \right) \Big\vert =
\partial_{\bar z}\left( D_{\pm}A^{\mp} \right) \Big\vert  \,.
\ee
We therefore identify
\be
\pm \overline{T}^{\pm}_{3\over 2} &=& 
\overline{D}_{\mp}\overline{X} \Big\vert 
= \pm \partial_{\bar z} \varphi^{\pm} \bar \psi^{\pm} \,, \qquad 
\pm \overline{\Theta}^{\pm}_{-{1\over 2}}=D_{\pm}A^{\mp} \Big\vert 
= \mp g \sin \varphi^{\pm} \psi^{\mp} \,.
\ee
Moreover,
\be
\partial_{z}\left( \overline{D}_{-}\overline{D}_{+}\overline{X} \right) 
\Big\vert =
\partial_{\bar z}\left( \overline{D}_{-}D_{-}A^{+} \right) \Big\vert \,,
\ee
and we therefore identify
\be
-\overline{T}_{2} = \overline{D}_{-}\overline{D}_{+}\overline{X} \Big\vert = 
-\partial_{\bar z} \varphi^{+}\partial_{\bar z}\varphi^{-} 
+ \bar \psi^{-} \partial_{\bar z} \bar \psi^{+} 
\,, \qquad
\overline{\Theta}_{0} = \Theta_{0} \,.
\ee

Consider now the quantity \cite{KU}
\be 
X = D_{+} \phi^{+} D_{-} \phi^{-} \left[ (\partial_{z}\phi^{+})^{2} +
(\partial_{z}\phi^{-})^{2} \right] - 2 \partial_{z} D_{+} \phi^{+}
\partial_{z} D_{-} \phi^{-} \,.
\ee
It obeys the conservation equations (\ref{superconservation}) with
\be
A^{\pm}= \mp g \cos \phi^{\mp}\left[ 3(\partial_{z}\phi^{\pm})^{2} +
(\partial_{z}\phi^{\mp})^{2} \right] 
- 2 g \sin \phi^{\mp}D_{+} \phi^{+} D_{-} \phi^{-} \partial_{z} 
\phi^{\pm} \,.
\ee 
It follows that
\be
T_{4} &=& D_{-}D_{+}X \Big\vert = 
-(\partial_{z} \varphi^{+})^{3}\partial_{z}\varphi^{-} 
-(\partial_{z} \varphi^{-})^{3}\partial_{z}\varphi^{+} \non  \\
&+& \psi^{-} \partial_{z} \psi^{+} \left[ 
(\partial_{z} \varphi^{+})^{2} + 3 (\partial_{z} \varphi^{-})^{2} 
\right] + 2 \psi^{+} \partial_{z} \psi^{-} (\partial_{z} \varphi^{+})^{2} 
\non  \\
&+& 2 \psi^{-} \psi^{+} \partial_{z} \varphi^{+} \partial_{z}^{2} \varphi^{+}
+ 2 \partial_{z}^{2} \varphi^{+} \partial_{z}^{2} \varphi^{-} 
- 2 \partial_{z} \psi^{-} \partial_{z}^{2} \psi^{+}  \,, \non \\
\Theta_{2} &=& \overline{D}_{-}D_{-}A^{+} \Big\vert =
g^{2} \sin \varphi^{-} \sin \varphi^{+} \left[ 
(\partial_{z} \varphi^{+})^{2} + (\partial_{z} \varphi^{-})^{2} \right] 
\non \\
&-&2 g^{2} \cos \varphi^{-} \cos \varphi^{+} \partial_{z} \varphi^{-}
\partial_{z} \varphi^{+} + 
g \cos \varphi^{-} \bar \psi^{+} \psi^{+} \left[ 
(\partial_{z} \varphi^{+})^{2} + (\partial_{z} \varphi^{-})^{2} \right]
\non \\
&+& 2 g \sin \varphi^{-} \partial_{z} \varphi^{-} \left(
-\psi^{+} \partial_{z} \bar\psi^{+} + \bar\psi^{+}\partial_{z}\psi^{+}
\right) 
- 2g \cos \varphi^{-} \partial_{z} \bar\psi^{+} \partial_{z}\psi^{+}
\,. \label{higher1}
\ee
Similarly, using 
\be 
\overline{X} = \overline{D}_{+} \phi^{+} \overline{D}_{-} \phi^{-} 
\left[ (\partial_{\bar z}\phi^{+})^{2} +
(\partial_{\bar z}\phi^{-})^{2} \right] 
- 2 \partial_{\bar z} \overline{D}_{+} \phi^{+}
\partial_{\bar z} \overline{D}_{-} \phi^{-} \,,
\ee
we obtain
\be
\overline{T}_{4} &=& \overline{D}_{-}\overline{D}_{+}\overline{X} 
\Big\vert = 
-(\partial_{\bar z} \varphi^{+})^{3}\partial_{\bar z}\varphi^{-} 
-(\partial_{\bar z} \varphi^{-})^{3}\partial_{\bar z}\varphi^{+} \non  \\
&+& \bar \psi^{-} \partial_{\bar z} \bar \psi^{+} \left[ 
(\partial_{\bar z} \varphi^{+})^{2} + 3 (\partial_{\bar z} \varphi^{-})^{2} 
\right] 
+ 2 \bar \psi^{+} \partial_{\bar z} \bar \psi^{-} 
(\partial_{\bar z} \varphi^{+})^{2} 
\non  \\
&+& 2 \bar \psi^{-} \bar \psi^{+} \partial_{\bar z} \varphi^{+} 
\partial_{\bar z}^{2} \varphi^{+}
+ 2 \partial_{\bar z}^{2} \varphi^{+} \partial_{\bar z}^{2} \varphi^{-} 
- 2 \partial_{\bar z} \bar \psi^{-} \partial_{\bar z}^{2} \bar \psi^{+}  
\,, \non \\
\overline{\Theta}_{2} &=& -\overline{D}_{-}D_{-}\overline{A}^{+} \Big\vert =
g^{2} \sin \varphi^{-} \sin \varphi^{+} \left[ 
(\partial_{\bar z} \varphi^{+})^{2} 
+ (\partial_{\bar z} \varphi^{-})^{2} \right] 
\non \\
&-&2 g^{2} \cos \varphi^{-} \cos \varphi^{+} \partial_{\bar z} \varphi^{-}
\partial_{\bar z} \varphi^{+} + 
g \cos \varphi^{-} \bar \psi^{+} \psi^{+} \left[ 
(\partial_{\bar z} \varphi^{+})^{2} 
+ (\partial_{\bar z} \varphi^{-})^{2} \right]
\non \\
&+& 2 g \sin \varphi^{-} \partial_{\bar z} \varphi^{-} \left(
- \psi^{+} \partial_{\bar z} \bar \psi^{+} 
+ \bar\psi^{+}\partial_{\bar z}\psi^{+} \right) 
- 2g \cos \varphi^{-} \partial_{\bar z} \bar\psi^{+} 
\partial_{\bar z}\psi^{+} \,. \label{higher2}
\ee

\end{document}